\documentclass[aps,pre,twocolumn,floats,showpacs]{revtex4}
\usepackage{epsfig}
\usepackage{color}
\usepackage{bm}
\usepackage{latexsym}

\begin{document}
\newcommand{\hide}[1]{}
\newcommand{\tbox}[1]{\mbox{\tiny #1}}
\newcommand{\half}{\mbox{\small $\frac{1}{2}$}}
\newcommand{\sinc}{\mbox{sinc}}
\newcommand{\const}{\mbox{const}}
\newcommand{\trc}{\mbox{trace}}
\newcommand{\intt}{\int\!\!\!\!\int }
\newcommand{\ointt}{\int\!\!\!\!\int\!\!\!\!\!\circ\ }
\newcommand{\eexp}{\mbox{e}^}
\newcommand{\bra}{\left\langle}
\newcommand{\ket}{\right\rangle}
\newcommand{\EPS} {\mbox{\LARGE $\epsilon$}}
\newcommand{\ar}{\mathsf r}
\newcommand{\im}{\mbox{Im}}
\newcommand{\re}{\mbox{Re}}
\newcommand{\bmsf}[1]{\bm{\mathsf{#1}}}
\newcommand{\mpg}[2][1.0\hsize]{\begin{minipage}[b]{#1}{#2}\end{minipage}}

\title{Scattering and transport properties of tight-binding random networks}

\author{A. J. Mart\'inez-Mendoza,$^{1,2}$ A. Alcazar-L\'opez,$^1$ and J. A. M\'endez-Berm\'udez$^1$}
\affiliation{$^1$Instituto de F\'{\i}sica, Benem\'erita Universidad Aut\'onoma de Puebla,
Apartado Postal J-48, Puebla 72570, Mexico \\ 
$^2$Elm\'eleti Fizika Tansz\'ek, Fizikai Int\'ezet, Budapesti M\H uszaki \'es Gazdas\'agtudom\'anyi Egyetem, H-1521 Budapest, Hungary}

\date{\today}

\begin{abstract}
We study numerically scattering and transport statistical properties 
of tight-binding random networks characterized by the number of nodes 
$N$ and the average connectivity $\alpha$. 
We use a scattering approach to electronic transport and concentrate on 
the case of a small number of single-channel attached leads. 
We observe a smooth crossover from insulating to
metallic behavior in the average scattering matrix elements $\bra |S_{mn}|^2 \ket$, 
the conductance probability distribution $w(T)$, the average conductance $\bra T \ket$, 
the shot noise power $P$, and the elastic enhancement factor $F$ by varying $\alpha$ 
from small ($\alpha \to 0$) to large ($\alpha \to 1$) values. 
We also show that all these quantities are invariant for fixed 
$\xi=\alpha N$.
Moreover, we proposes a heuristic and universal relation between 
$\bra |S_{mn}|^2 \ket$, $\bra T \ket$, and $P$ and the disorder parameter $\xi$.
\end{abstract}


\maketitle

%%%%%%%%%%%%%%%%%%%%%%%%%%%%     Model     %%%%%%%%%%%%%%%%%%%%%%%%%%%%%%%%%%%%%

\section{Introduction and model}

During the last three decades there has been an increasing number of papers
devoted to the study of random graphs and complex networks, in view of the fact 
that they describe systems in many knowledge areas: from maths and physics to 
finance and social sciences, passing through biology and chemistry \cite{BA99,S01,AB02,B13}. 
In particular, some of those works report studies of spectral and eigenfunction 
properties of complex networks; see for example Refs.~\cite{DR93a,ZX00,DR93b,GGS05,SKHB05,BJ07a,BJ07b,F02,DGMS03,GT06a,GT06b,EK09,HS11,AMM}.
That is, since complex networks composed by nodes and the bonds joining them can
be represented by sparse matrices, it is quite natural to ask about the spectral 
and eigenfunction properties of such {\it adjacency} matrices.
Then, in fact, studies originally motivated on physical systems represented by 
Hamiltonian sparse random matrices \cite{RB88,RD90,FM91,EE92,JMR01} can be directly 
applied to complex networks.

In contrast to the numerous works devoted to study spectral and 
eigenfunction properties of complex netwoks, to our knowledge, 
just a few focus on some of their scattering and transport properties 
\cite{MPB07a,MPB07b,XLL08,SRS10,PW13}. 
So, in the present work we study numerically several statistical properties
of the scattering matrix and the electronic transport across disordered
tight-binding networks described by sparse real symmetric matrices.
We stress that we use a scattering approach to electronic transport; see
for example \cite{MK04}. In addition, we concentrate on the case of a small 
number of attached leads (or terminals), each of them supporting one open channel.
We also note that tight-binding complex networks have also been studied in
Refs.~\cite{DR93a,ZX00,GT06a,GT06b}.

The tight-binding random networks we shall study here are described by the 
tight-binding Hamiltonian 
\begin{equation}
\label{TBH}
H = \sum^N_{n=1} h_{nn} | n \rangle \langle n| +
\sum^N_{n=1} \sum^N_{m=1} h_{nm} \left( | n \rangle \langle m| + |m \rangle \langle n| \right) \ ,
\end{equation}
where $N$ is the number of nodes or vertexes in the network, $h_{nn}$ are on-site 
potentials and $h_{nm}$ are the hopping integrals between sites $n$ and $m$.
Then we choose $H$ to be a member of an ensemble of $N\times N$ sparse 
real symmetric matrices whose nonvanishing elements are statistically independent random 
variables drawn from a normal distribution with zero mean $\bra h_{nm} \ket=0$ 
and variance $\bra |h_{nm}|^2 \ket=(1+\delta_{nm})/2$. As in Refs.~\cite{AMM,JMR01},
here we define the sparsity of $H$, $\alpha$, as the fraction of the $N(N-1)/2$ nonvanishing 
off-diagonal matrix elements. I.e., $\alpha$ is the network average connectivity.
Thus, our random network model corresponds to an ensemble of adjacency 
matrices of Erd\H{o}s-R\'enyi--type random graphs \cite{ER59,AB02,note1}.

Notice that with the prescription given above our network model displays {\it 
maximal disorder} since averaging over the network ensemble implies average over 
connectivity and over on-site potentials and hopping integrals.
With this averaging procedure we get rid off any individual network characteristic
(such as {\it scars} \cite{SK03} which in turn produce topological resonances 
\cite{GSS13}) that may lead to deviations from random matrix theory (RMT) predictions
which we use as a reference. I.e., we choose this network model to retrieve well known 
random matrices in the appropriate limits: a diagonal random matrix is obtained 
for $\alpha=0$ when the nodes in the network are isolated, while a member of the 
Gaussian Orthogonal Ensemble (GOE) is recovered for $\alpha=1$ when the network 
is fully connected.

However, it is important to add that the {\it maximal disorder} we consider is not 
necessary for a graph/network to exhibit universal RMT behavior. In fact:
(i) It is well known that tight-binding cubic lattices with on-site disorder (known
as the three-dimensional Anderson model \cite{3DAM}), forming networks with fixed 
regular connectivity having a very dilute Hamiltonian matrix, show RMT behavior in 
the {\it metallic phase} (see for example Refs.~\cite{metallic1,metallic2}). 
(ii) It has been demonstrated numerically and theoretically that graphs 
with fixed connectivity show spectral \cite{spectral,TS01} and scattering 
\cite{PW13,scattering} universal properties corresponding to RMT predictions, 
where in this case the disorder is introduced either by choosing random bond lengths
\cite{spectral,PW13,scattering} (which is a parameter not persent in our network model) 
or by randomizing the vertex-scattering matrices \cite{TS01} 
(somehow equivalent to consider random on-site potentials).
Moreover, some of the RMT properties of quantum graphs have already been tested 
experimentally by the use of small ensembles of small microwave networks with fixed 
connectivity \cite{Sirko}.
(iii) Complex networks having specific topological properties (such as small-world 
and scale-free networks, among others), where randomness is 
applied only to the connectivity, show signatures of RMT behavior in their 
spectral and eigenfunction properties \cite{BJ07a,GT06a,MPB07a}.

The organization of this paper is as follows.
In the next section we define the scattering setup as well as the scattering 
quantities under investigation and provide the corresponding analytical 
predictions from random scattering-matrix theory for systems with
time-reversal symmetry. These analytical results will be used as a
reference along the paper. In Section III we analyze the average
scattering matrix elements $\bra |S_{mn}|^2 \ket$, the conductance probability distribution 
$w(T)$, the average conductance $\bra T \ket$, the shot noise power $P$, and 
the elastic enhancement factor $F$ for 
tight-binding networks as a function of $N$ and $\alpha$. 
We show that all scattering and transport quantities listed above are invariant 
for fixed $\xi$. 
Moreover, we propose a heuristic and universal relation between 
$\bra |S_{mn}|^2 \ket$, $\bra T \ket$, and $P$ and the disorder parameter $\xi$.
Finally, Section IV is left for conclusions.

\section{The scattering setup and RMT predictions}

We open the isolated samples, defined above by the tight-binding random network 
model, by attaching $2M$ semi-infinite single channel leads. Each lead is 
described by the one-dimensional semi-infinite tight-binding Hamiltonian
\begin{equation}
\label{leads}
H_{\tbox{lead}}=\sum^{-\infty}_{n=1} (| n \rangle \langle n+1| + |n+1 \rangle \langle n|) \ .
\end{equation}
Using standard methods one can write the scattering
matrix ($S$-matrix) in the form \cite{MW69} 
\begin{equation}
\label{smatrix}
S(E) =
\left(
\begin{array}{cc}
r & t'   \\
t & r'
\end{array}
\right)
={\bf 1}-2i \sin (k)\, {\cal W}^{\,T} (E-{\cal H}_{\rm eff})^{-1} {\cal W} \ ,
\end{equation}
where $t$, $t'$, $r$, and $r'$ are $M\times M$ transmission and reflection
matrices; ${\bf 1}$ is the $2M\times 2M$ unit matrix, $k=\arccos(E/2)$ is
the wave vector supported in the leads, and ${\cal H}_{\rm eff}$ is
an effective non-hermitian Hamiltonian given by
\begin{equation}
\label{Heff}
{\mathcal{H}}_{\rm eff}=H- e^{ik} {\cal W}{\cal W}^{\,T} \ .
\end{equation}
Here, ${\cal W}$ is an $N\times 2M$ matrix that specifies the positions
of the attached leads to the network. However, in the random network model
we are studying here all nodes are equivalent; so, we attach the $2M$ 
leads to $2M$ randomly chosen nodes. The elements of ${\cal W}$ are equal 
to zero or 
$\epsilon$, where $\epsilon$ is the coupling strength. Moreover, assuming
that the wave vector $k$ do not change significantly in the center of the
band, we set $E=0$ and neglect the energy dependence of
${\mathcal{H}}_{\rm eff}$ and $S$.

Since in the limit $\alpha=1$ the random network model reproduces the 
GOE, in that limit we expect the 
statistics of the scattering matrix, Eq.~(\ref{smatrix}), to be determined 
by the Circular Orthogonal Ensemble (COE) which is the appropriate scattering matrix 
ensemble for {\it internal} systems $H$ with time reversal symmetry. Thus, 
below, we provide the 
statistical results for the $S$-matrix and the transport quantities 
to be analyzed in the following sections, assuming the orthogonal symmetry.
In all cases, we also assume the absence of direct processes (also known as 
perfect coupling condition), i.e., $\langle S \rangle=0$.

We start with the average of the $S$-matrix elements.
It is known that
\begin{equation}
\label{Saa}
\bra |S_{mn}|^2 \ket_{\tbox{COE}} = \frac{1+\delta_{mn}}{2M+1} \ ,
\end{equation}
where $\bra \cdot \ket$ means ensemble average over the COE.

Within a scattering approach to the electronic transport, once the
scattering matrix is known one can compute the dimensionless
conductance \cite{Landauer}
\[
T={\mbox{Tr}}(tt^\dagger)=\sum_m\sum_n|t_{mn}|^2
\]
and its distribution $w(T)$.
For $M=1$, i.e. considering two single-channel leads attached to
the network, $w(T)$ is given by
\begin{equation}
\label{wofTM1}
w(T)_{\tbox{COE}} = \frac{1}{2\sqrt{T}} \ ,
\end{equation}
while for $M=2$,
\begin{equation}
w(T)_{\tbox{COE}} = \left\{
\label{wofTM2}
\begin{array}{ll}
3T/2 \ , & 0<T<1  \\
3\left( T-2\sqrt{T-1}\right)/2 \ , & 1<T<2 \\
\end{array} \right. \ . \\
\end{equation}
For arbitrary $M$, the prediction for the average value of $T$ is
\begin{equation}
\label{avT}
\bra T \ket_{\tbox{COE}} = \frac{M}{2}-\frac{M}{2(2M+1)} \ .
\end{equation}
For the derivation of the expressions above see for example 
Ref.~\cite{MK04}.
A related transport quantity is the shot noise power
\[
P = \bra {\mbox{Tr}}(tt^\dagger - tt^\dagger tt^\dagger) \ket \ ,
\]
which as a function of $M$ reads \cite{evgeny}
\begin{equation}
\label{P}
P_{\tbox{COE}} = \frac{M(M+1)^2}{2(2M+1)(2M+3)} \ .
\end{equation}

Another scattering quantity of interest that measures {\it cross sections} 
fluctuations is the elastic enhancement factor \cite{EF}
\begin{equation}
\label{F}
F = \frac{\bra |S_{mm}|^2 \ket}{\bra |S_{mn}|^2 \ket} \ ,
\end{equation}
that in the RMT limit becomes
\begin{equation}
\label{FCOE}
F_{\tbox{COE}}=2 \ .
\end{equation}

In the following sections we focus on $\bra |S_{mn}|^2 \ket$,
$\bra T \ket$, $P$, and $F$ for the tight-binding random network model.

\section{Results}

\begin{figure}[t]
\centerline{\includegraphics[width=7cm]{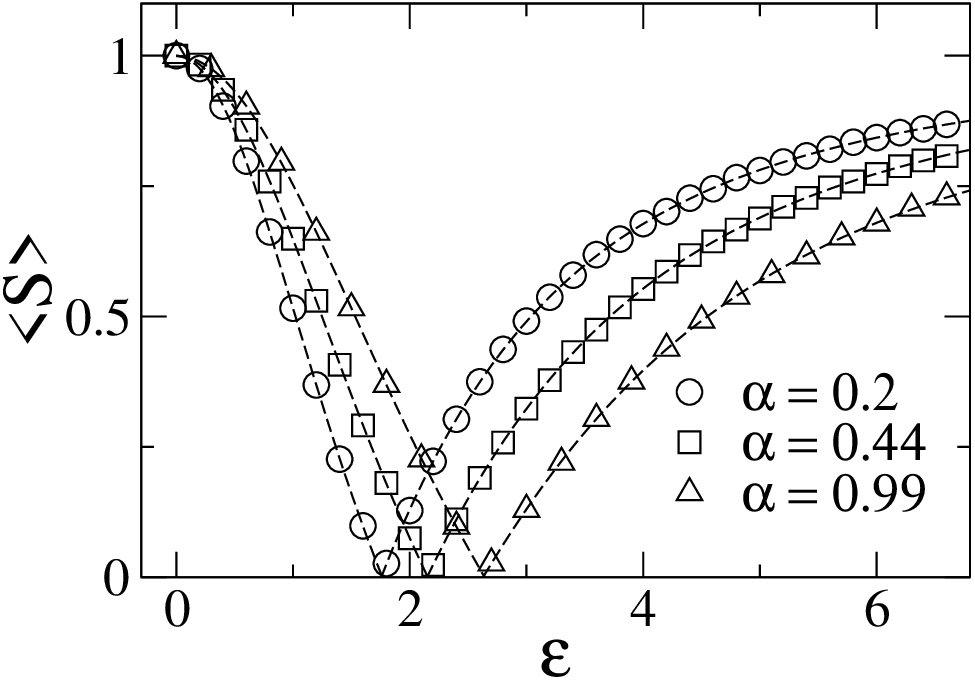}}
\caption{Average $S$-matrix, as defined in Eq.~(\ref{avS}), for tight-binding 
random networks having $N=50$ nodes as a function of the coupling strength 
$\epsilon$. We found $\epsilon_0\approx 1.76$, 2.15, and 2.63 for $\alpha=0.2$,
0.44, and 0.99, respectively. Dashed lines are fittings of Eq.~(\ref{Svse}) to
the data. Each point was computed by averaging over $10^6$ 
random network realizations.}
\label{Fig1}
\end{figure}

In all cases below we set the coupling strength $\epsilon$ such that 
\begin{equation}
\label{avS}
\bra S \ket \equiv \frac{1}{2M} \sum_{mn} | \bra S_{mn} \ket |
\end{equation}
is approximately zero in order to compare our results, in the limit 
$\alpha\to 1$, with the RMT predictions reviewed above, see 
Eqs.~(\ref{Saa}-\ref{P}) and (\ref{FCOE}). 
To find the perfect coupling condition we plot $\bra S \ket$ vs.~$\epsilon$ 
for fixed $N$ and $\alpha$ and look for the minimum.
As an example, in Fig.~\ref{Fig1} we plot $\bra S \ket$ vs.~$\epsilon$
for random networks having $N=50$ nodes with $\alpha=0.2$,
0.44, and 0.99. Notice that: For $\epsilon=0$, $\bra S \ket=1$; i.e., since there is no
coupling between the network and the leads, there is total reflection
of the waves incoming from the leads. While since for any $\epsilon>0$
the waves do interact with the random network, $\bra S \ket<1$. 

It is clear from Fig.~\ref{Fig1} that the curves $\bra S \ket$ vs.~$\epsilon$ behave
similarly. In fact we identify two regimes:
When $0<\epsilon<\epsilon_0$, $\bra S \ket$ decreases with $\epsilon$; while 
for $\epsilon>\epsilon_0$, $\bra S \ket$ increases with $\epsilon$.
Since $\epsilon_0$ is the coupling strength value at which $\bra S \ket\approx 0$,
we set $\epsilon=\epsilon_0$ to achieve the perfect coupling condition.

In addition, as in previous studies \cite{MMV09,AMV09}, here we found 
that the curves $\bra S \ket$ vs.~$\epsilon$ are well fitted by the expression
\begin{equation}
\label{Svse}
\bra S \ket = \frac{C_0}{1+(C_1\epsilon)^{\pm C_2}} - C_3 \ ,
\end{equation}
where $C_i$ are fitting constants and the plus and minus signs correspond 
to the regions $0<\epsilon<\epsilon_0$ and $\epsilon>\epsilon_0$, respectively. 
With the help of Eq.~(\ref{Svse}) we can find $\epsilon_0$ with a relatively
small number of data points.
Moreover, we heuristically found that
\begin{equation}
\epsilon_0 \approx (\alpha \cdot N)^{1/4} \ .
\end{equation}
Then, we use this prescription to compute $\epsilon_0$ which is the 
value for the coupling strength that we set in all the calculations below.

In the following, all quantities and histograms were computed by the use
of $10^6$ random network realizations for each combination of $N$ and $\alpha$. 

\subsection{Average scattering matrix elements}

\begin{figure}[t]
\centerline{\includegraphics[width=7cm]{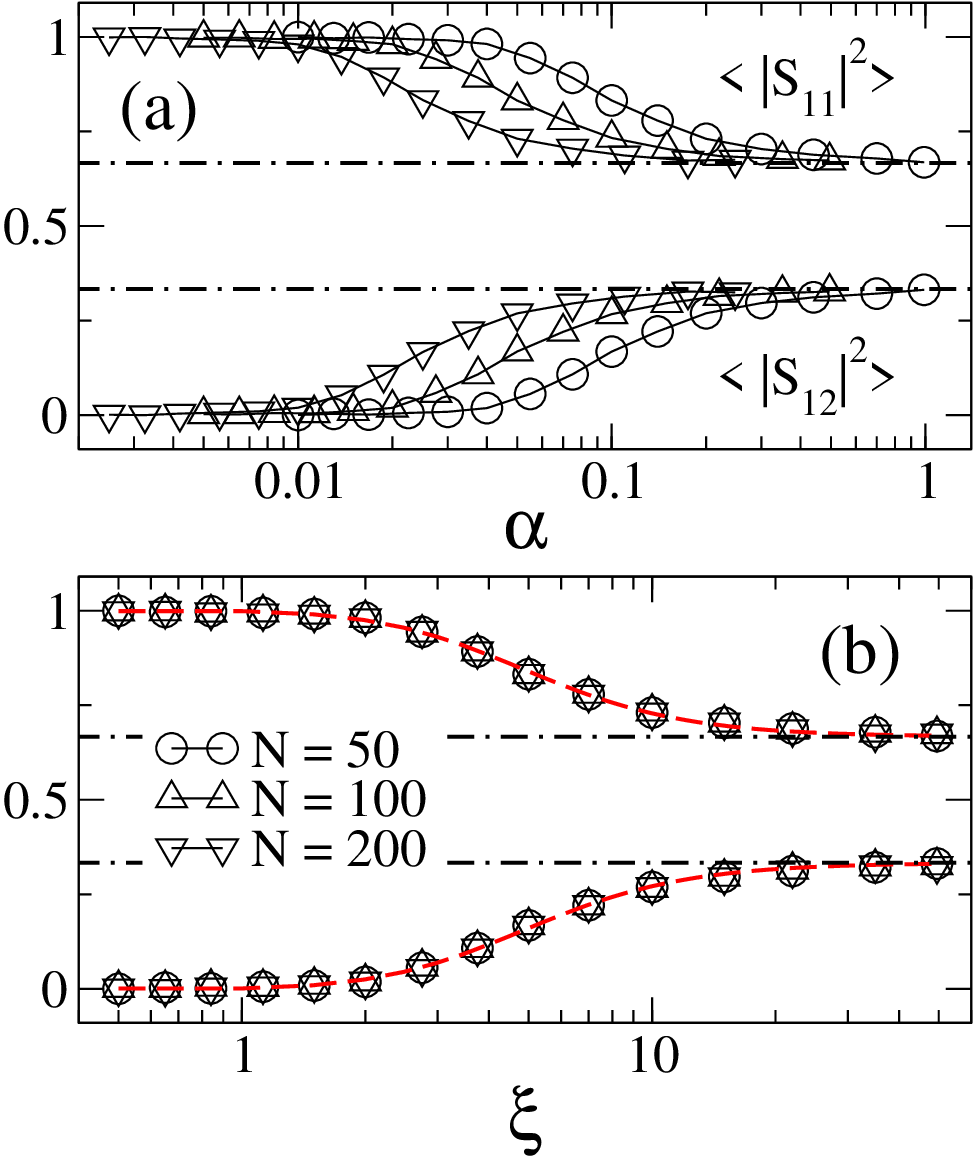}}
\caption{(Color online) Average $S$-matrix elements $\bra |S_{11}|^2 \ket$ and 
$\bra |S_{12}|^2 \ket$ for tight-binding random networks having $N$ nodes as 
a function of (a) $\alpha$ and (b) $\xi$, for $M=1$. 
The dot-dashed lines correspond to 2/3 and 1/3; the RMT prediction 
for $\bra |S_{11}|^2 \ket$ and $\bra |S_{12}|^2 \ket$, respectively, 
given by Eq.~(\ref{Saa}). Red dashed lines in (b) are 
Eqs.~(\ref{S11x}) and (\ref{S12x}) with $\delta \approx 0.198$.
Error bars in this and the following figures are not shown since they are much 
smaller than symbol size.}
\label{Fig2}
\end{figure}

First we consider the case $M=1$, where the $S$-matrix is a $2\times 2$ 
matrix. In Fig.~\ref{Fig2}(a) we plot the ensemble average of the elements 
$|S_{11}|^2$ (average reflexion) and $|S_{12}|^2$ (average transmission) 
as a function of the connectivity $\alpha$ for three different network sizes. 
The COE limit, Eq.~(\ref{Saa}), expected for $\alpha\to 1$ is 
also plotted (dot-dashed lines) as reference. 
Notice that for all three network sizes the behavior is similar: there is a 
strong $\alpha$-dependence of the average $S$-matrix elements driving 
the random network from a localized or insulating regime [$\bra |S_{11}|^2 \ket \approx 1$ and
$\bra |S_{12}|^2 \ket \approx 0$; i.e., the average conductance is close 
to zero] for $\alpha\to 0$, to a delocalized or metallic regime 
[$\bra |S_{11}|^2 \ket \approx 2/3$ and $\bra |S_{12}|^2 \ket \approx 1/3$; 
i.e., RMT results are already recovered] for $\alpha \to 1$.
Moreover, the curves $\bra |S_{mn}|^2 \ket$ vs.~$\alpha$ are displaced along
the $\alpha$-axis: the larger the network size $N$ the smaller the value of
$\alpha$ needed to approach the COE limit. 

We now recall that the parameter  
\begin{equation}
\xi \equiv \alpha \times N 
\label{xi}
\end{equation}
was shown to fix (i) spectral properties of sparse random matrices \cite{JMR01},
(ii) the percolation transition of Erd\H{o}s-R\'enyi random graphs, see for 
example Ref.~\cite{AB02}, where $\xi$ has the name of average degree;
and (iii) the nearest-neighbor energy level spacing distribution and the entropic 
eigenfunction localization length of sparse random matrices \cite{AMM}. 
So, it make sense to explore the dependence of $\bra |S_{mn}|^2 \ket$ on $\xi$.
Then, in Fig.~\ref{Fig2}(b) we plot again $\bra |S_{11}|^2 \ket$ and 
$\bra |S_{12}|^2 \ket$ but now as a function of
$\xi$. We observe that curves for different $N$ now fall on top of a universal
curve. 

\begin{figure}[t]
\centerline{\includegraphics[width=7cm]{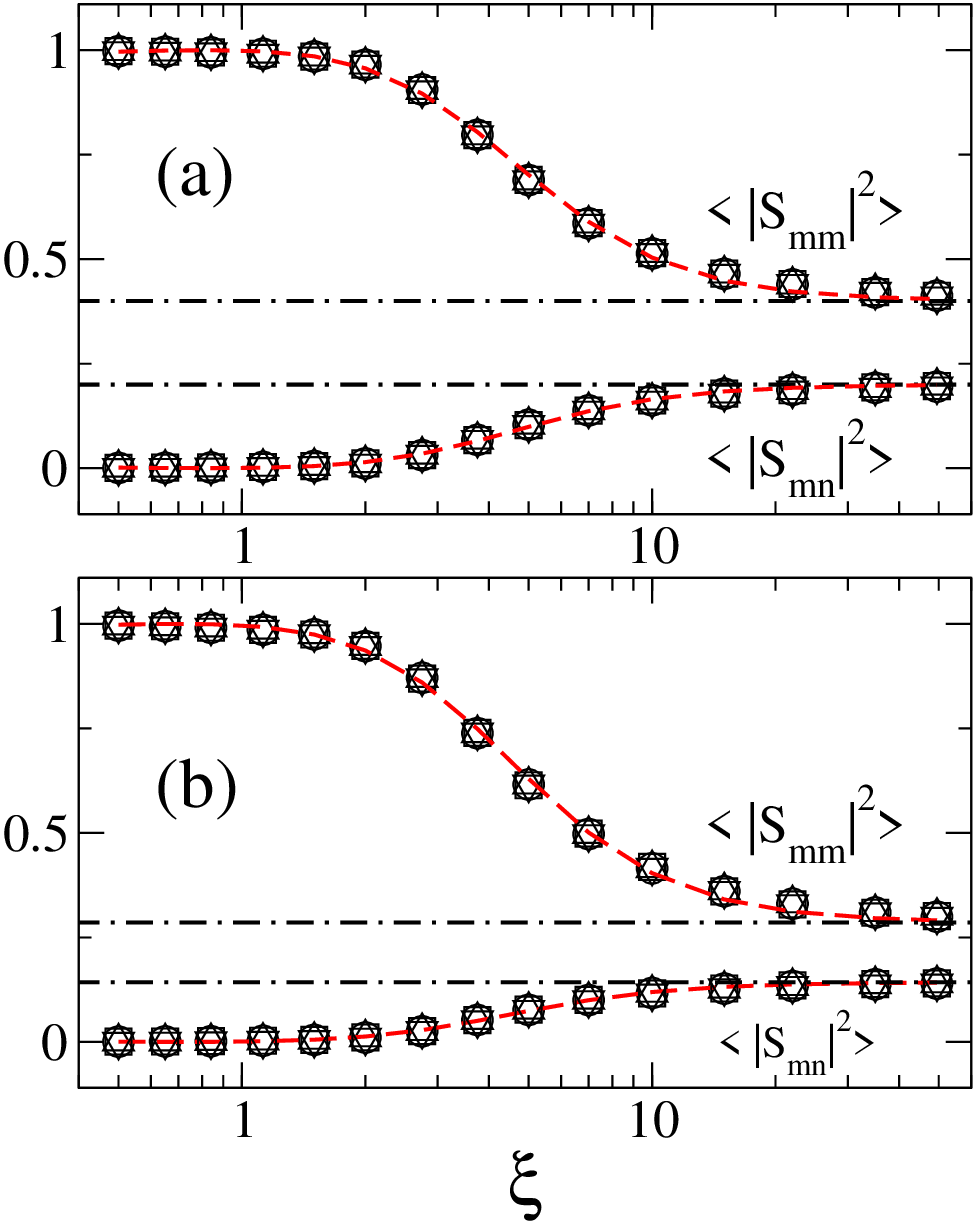}}
\caption{(Color online) Average $S$-matrix elements $\bra |S_{mm}|^2 \ket$ 
[with $mm=11$, 22, 33, and 44] and $\bra |S_{mn}|^2 \ket$ [with $mn=12$, 
23, 34, and 41] for tight-binding random networks having $N=200$ 
nodes as a function of $\xi$ for (a) $M=2$ and (b) $M=3$. 
The dot-dashed lines correspond to the RMT prediction for 
$\bra |S_{mm}|^2 \ket$ and $\bra |S_{mn}|^2 \ket$; see Eq.~(\ref{Saa}). 
Red dashed lines are Eqs.~(\ref{Smmx}) and (\ref{Smnx}) with 
(a) $\delta \approx 0.237$ and (b) $\delta \approx 0.242$.}
\label{Fig3}
\end{figure}

Moreover, we have found that the universal behavior of $\bra |S_{11}|^2 \ket$ 
and $\bra |S_{12}|^2 \ket$, as a function of $\xi$, is well described by
\begin{eqnarray}
\label{S11x}
\bra |S_{11}|^2 \ket & = & 1 - \bra |S_{12}|^2 \ket \ , \\
\label{S12x}
\bra |S_{12}|^2 \ket & = & \frac{1}{3} 
\left[ \frac{1}{1+(\delta \xi)^{-2}} \right] \ ,
\end{eqnarray}
where $\delta$ is a fitting parameter. Eq.~(\ref{S11x}) is a
consequence of the unitarity of the scattering matrix, 
$SS^\dagger =\mathbf 1$, while the factor 1/3 in Eq.~(\ref{S12x})
comes from Eq.~(\ref{Saa}) with $M=1$.
In Fig.~\ref{Fig2}(b) we also include Eqs.~(\ref{S11x}) and (\ref{S12x})
(red dashed lines) and observe that they reproduce very well the corresponding 
numerical results. 
In fact, we have to add that Eqs.~(\ref{S11x}) and 
(\ref{S12x}) also work well for other random 
matrix models showing a metal-insulator phase transition \cite{AMV09}.

For $M>1$ we observe the same scenario as for $M=1$: All $S$-matrix elements
suffer a localization-delocalization transition as a function of $\xi$.
See Fig.~\ref{Fig3} where we plot some of the average $S$-matrix elements
for $M=2$ and 3. Moreover, we were able to generalize Eqs.~(\ref{S11x}) and 
(\ref{S12x}) to any $M$ as
\begin{eqnarray}
\label{Smmx}
\bra |S_{mm}|^2 \ket & = & 1 - (2M-1)\bra |S_{mn}|^2 \ket \ , \\
\label{Smnx}
\bra |S_{mn}|^2 \ket & = & \bra |S_{mn}|^2 \ket_{\tbox{COE}} 
\left[ \frac{1}{1+(\delta \xi)^{-2}} \right] \ .
\end{eqnarray}
Then, in Fig.~\ref{Fig3} we also plot Eqs.~(\ref{Smmx}) and (\ref{Smnx}) 
and observe very good correspondence with the numerical data.
We also note that the fitting parameter $\delta$ slightly depends on $M$.

Finally we want to remark that concerning $\bra |S_{mn}|^2 \ket$, the RMT
limit, expected for $\alpha\to 1$ or $\xi\to N$, is already recovered for
$\xi\ge 30$.

\begin{figure}[t]
\centerline{\includegraphics[width=7.5cm]{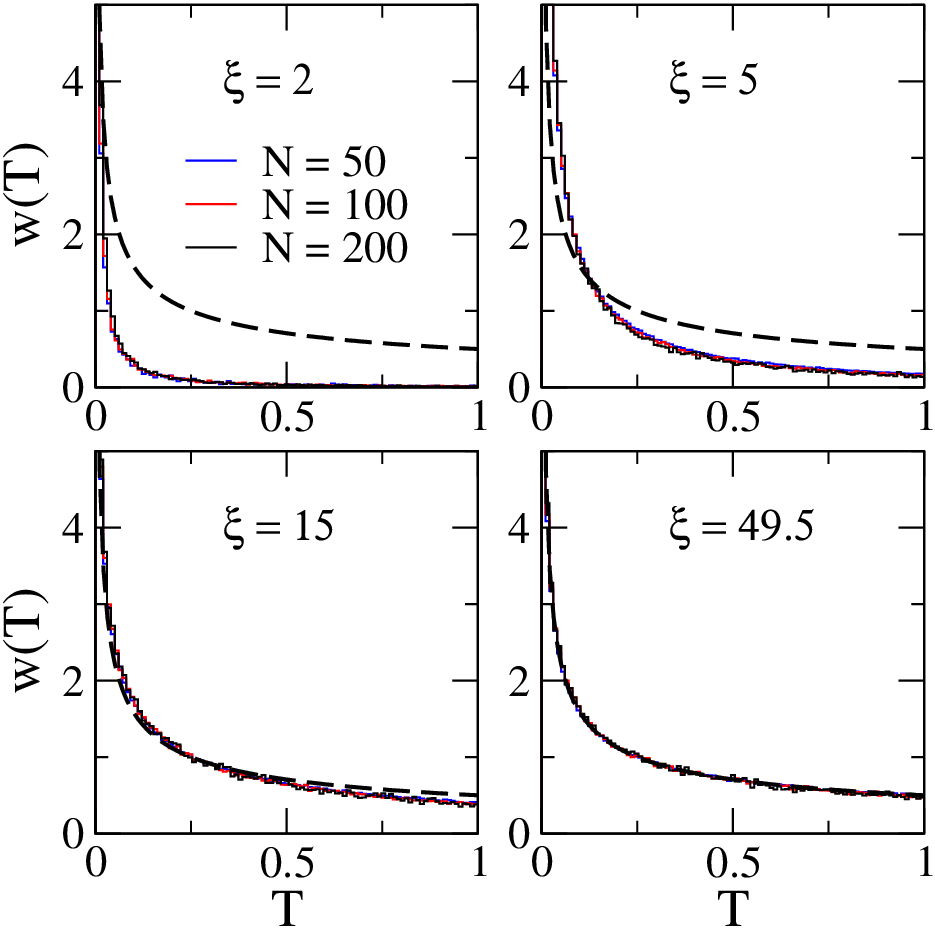}}
\caption{(Color online) Conductance probability distribution $w(T)$ 
for tight-binding random networks having $N$ nodes, in the case $M=1$, for some
values of $\xi$. Dashed lines are $w(T)_{\tbox{COE}}$; the RMT 
prediction for $w(T)$ given by Eq.~(\ref{wofTM1}).}
\label{Fig4}
\end{figure}
\begin{figure}[t]
\centerline{\includegraphics[width=7.5cm]{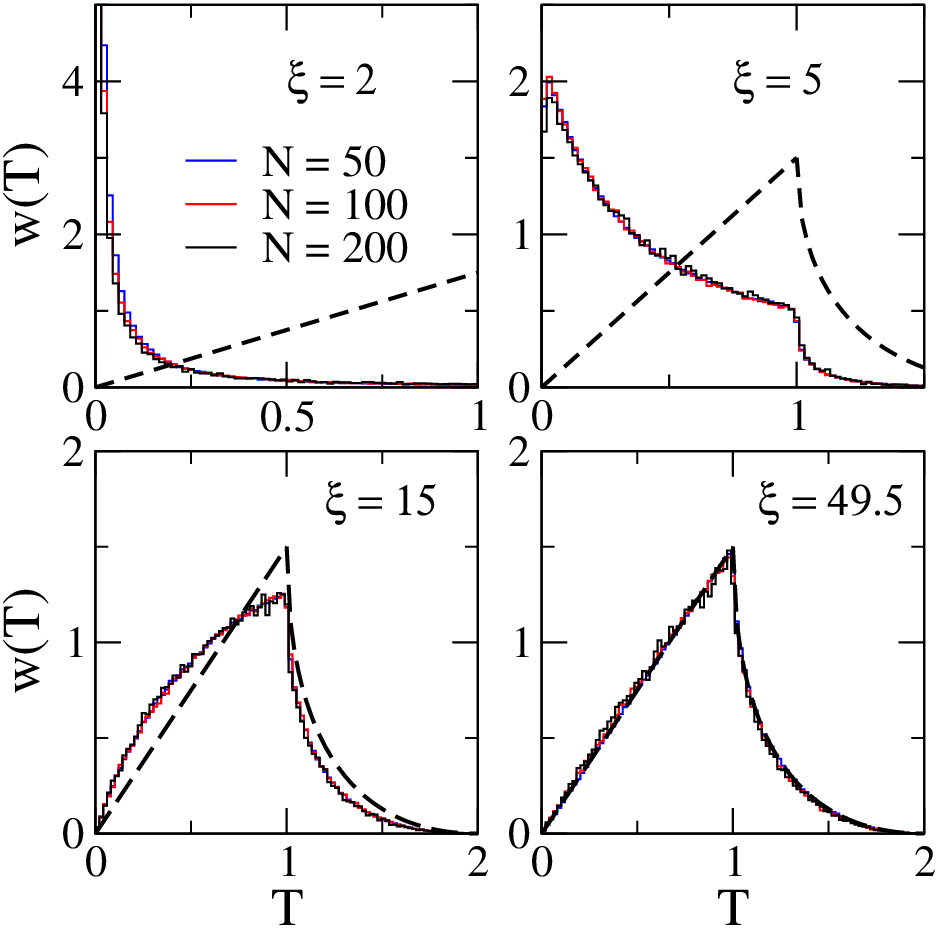}}
\caption{(Color online) Conductance probability distribution $w(T)$ 
for tight-binding random networks having $N$ nodes, in the case $M=2$, for some
values of $\xi$. Dashed lines are $w(T)_{\tbox{COE}}$; the RMT 
prediction for $w(T)$ given by Eq.~(\ref{wofTM2}).}
\label{Fig5}
\end{figure}

\subsection{Conductance and shot noise power}

Now we turn to the conductance statistics.
In Figs.~\ref{Fig4} and \ref{Fig5} we present conductance probability 
distributions $w(T)$ for $M=1$ and $M=2$, respectively. In both cases 
we include the corresponding RMT predictions. We report histograms for 
four values of $\xi$ and three network sizes.
From these figures, it is clear that $w(T)$ is invariant once $\xi$ is
fixed; i.e., once $\xi$ is set to a given value, $w(T)$ does not depend
on the size of the network. 
We also recall that in the limit $\alpha\to 1$, $w(T)$ is expected to 
approach the RMT predictions of Eqs.~(\ref{wofTM1}) and (\ref{wofTM2}).
However, we observe that $w(T)$ is already well described by 
$w(T)_{\tbox{COE}}$ once $\xi\ge 30$.
We observe an equivalent scenario for $w(T)$ when $M>2$ (not shown here).

\begin{figure}[t]
\centerline{\includegraphics[width=8cm]{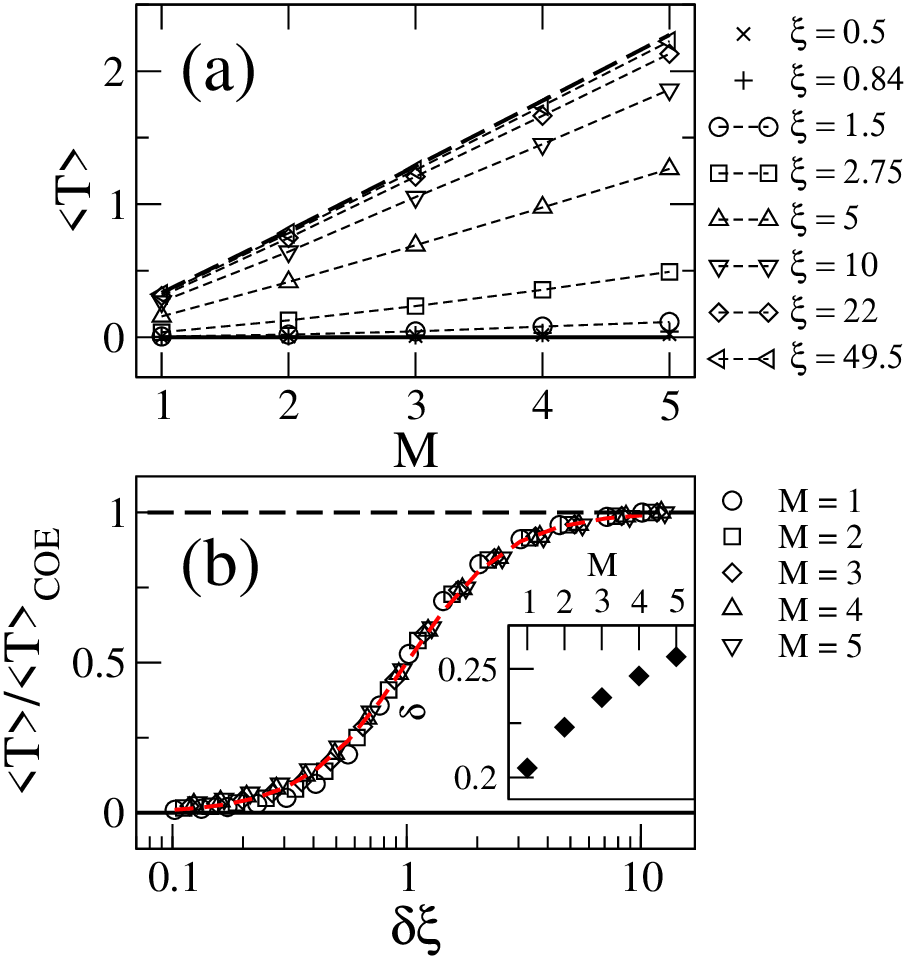}}
\caption{(Color online) (a) Average conductance $\bra T \ket$ as a function of $M$ 
for tight-binding random networks having $N=200$ nodes for several values 
of $\xi$. (b) $\bra T \ket/\bra T \ket_{\tbox{COE}}$ as a function of 
$\delta\xi$ for $M\in[1,5]$. Insert: $\delta$ versus $M$. $\delta$ 
is obtained from the fitting of Eq.~(\ref{Xxi}) to the $\bra T \ket$ 
vs.~$\xi$ data. Thick full lines correspond to $\bra T \ket = 0$.
Dashed lines are (a) the RMT prediction for $\bra T \ket$, given by 
Eq.~(\ref{avT}); and (b) one. The red dashed line in (b) on top of the 
data is Eq.~(\ref{Xxi}).}
\label{Fig6}
\end{figure}

\begin{figure}[t]
\centerline{\includegraphics[width=8cm]{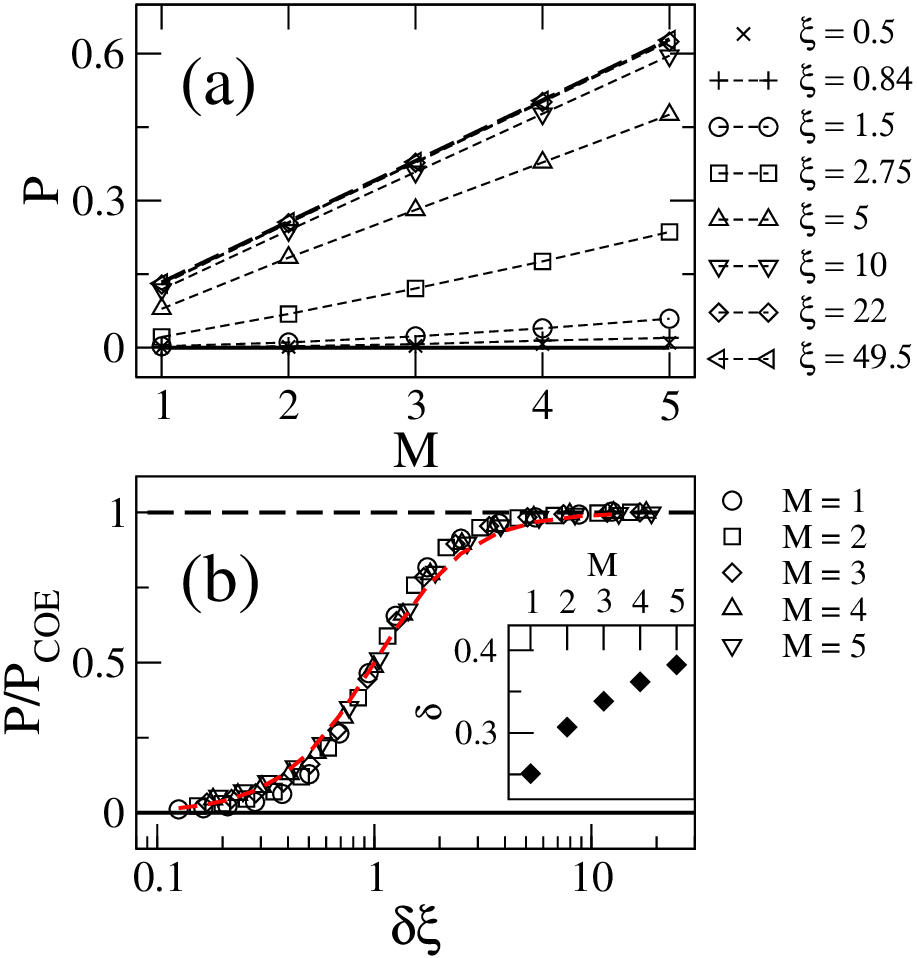}}
\caption{(Color online) (a) Shot noise power $P$ as a function of $M$ for tight-binding 
random networks having $N=200$ nodes for several values of $\xi$.
(b) $P/P_{\tbox{COE}}$ as a function of $\delta\xi$ for $M\in[1,5]$. 
Insert: $\delta$ versus $M$. $\delta$ is obtained from the fitting 
of Eq.~(\ref{Xxi}) to the $P$ vs.~$\xi$ data. Thick full lines 
correspond to $P = 0$. Dashed lines are (a) the RMT 
prediction for $P$, given by Eq.~(\ref{P}); and (b) one. The red dashed 
line in (b) on top of the data is Eq.~(\ref{Xxi}).}
\label{Fig7}
\end{figure}

We now increase further the number of attached leads.
Then, in Figs.~\ref{Fig6}(a) and \ref{Fig7}(a) we plot
the average conductance $\bra T \ket$ and the shot noise power $P$ for
tight-binding random networks having $N=200$ nodes, for several values
of $\xi$ with $M\in [1,5]$ (we recall that for $M=5$, ten single-channel 
leads are attached to the networks).
It is clear from these plots that changing $\xi$ from small ($\xi< 1$) to 
large ($\xi\gg 1$) values produces a transition from localized to 
delocalized behavior in the scattering properties of random notworks. 
That is, (i) for $\xi<0.5$, $\bra T \ket \approx 0$ and $P \approx 0$; 
and (ii) for $\xi\ge 30$, $\bra T \ket$ and $P$ are well given by the 
corresponding RMT predictions given by Eqs.~(\ref{avT}) and (\ref{P}), 
respectively.
Equivalent plots are obtained (not shown here) for other network sizes.

Moreover, we have observed that $\bra T \ket$ and $P$ as a function of $\xi$
behave (for all 
$M$) as $\bra |S_{mn}|^2\ket$ does. I.e., they show a universal behavior 
as a function of $\delta\xi$ that can be well described by
\begin{equation}
\label{Xxi}
X(\xi) = X_{\tbox{COE}} \left[ \frac{1}{1+(\delta\xi)^{-2}}
\right] \ ,
\end{equation}
where $X$ represents $\bra T \ket$ or $P$ and $\delta$ is the fitting 
parameter.
Then, in Figs.~\ref{Fig6}(b) and \ref{Fig7}(b) we plot $\bra T \ket$
and $P$ normalized to their respective COE average values, as a function 
of $\delta\xi$ for $M\in [1,5]$.
Notice that all curves for different $M$ fall on top of the universal 
curve given by Eq.~(\ref{Xxi}). 


\subsection{Enhancement factor}

\begin{figure}[t]
\centerline{\includegraphics[width=7cm]{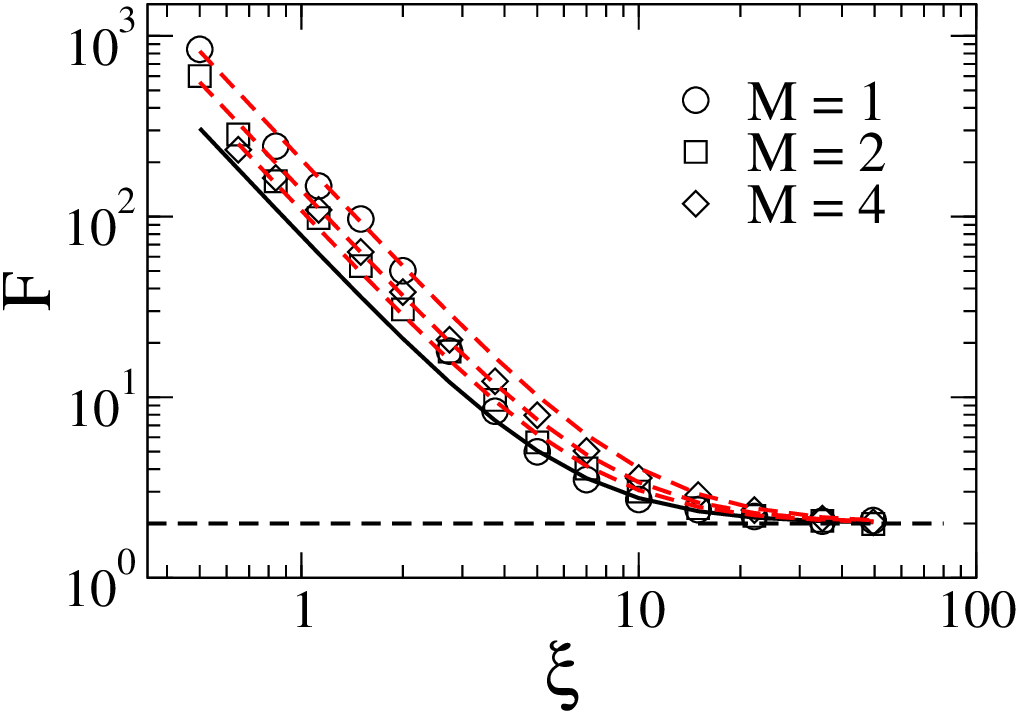}}
\caption{(Color online) Elastic enhancement factor $F$ as a function of 
$\xi$ for tight-binding random networks having $N=50$ nodes for $M=1$, 
2, and 4. Black full line is Eq.~(\ref{F1}) with $M=1$ and $\delta=0.198$.
Red dashed lines are fittings of Eq.~(\ref{F2}) to the data with $C=205$, 
138, and 106 for $M=1$, 2 and 4, respectively. The horizontal black dashed 
line corresponds to the RMT limit value of $F_{\tbox{COE}}=2$.}
\label{Fig8}
\end{figure}

Finally, in Fig.~\ref{Fig8} we plot the elastic enhancement factor $F$ as a function of 
$\xi$ for random networks with $N=50$ nodes for $M=1$, 2, and 4.
From this figure we observe that, for any $M$ (and also for any $N$, not shown 
here), $F$ decreases as a function of $\xi$ and approaches smoothly, for large 
$\xi$ ($\xi\to N$), the RMT limit value of $F_{\tbox{COE}}=2$. Also note that when 
$\xi\ll 1$, $F\propto \xi^{-2}$; which seems to be a signature of our random network model.

To have an analytic support for the observations made above, we substitute 
Eqs.~(\ref{Smmx}) and (\ref{Smnx}) into Eq.~(\ref{F}) to get the following estimation 
for $F$:
\begin{equation}
\label{F1}
F \approx (2M+1)(\delta\xi)^{-2} + 2 \ .
\end{equation}
Notice that Eq.~(\ref{F1}) reproduces properly the behavior of $F$ for small 
and large $\xi$: $F\propto \xi^{-2}$ and $F\to 2$, respectively.
Unfortunately, Eq.~(\ref{F1}) does not describe qualitatively the curves of
Fig.~\ref{Fig8}, see as example the black full line in this figure that
corresponds to Eq.~(\ref{F1}) with $M=1$. 
The reason of this discrepancy, as a detailed analysis shows, is that 
Eq.~(\ref{Smnx}) overestimates the magnitude of $\bra |S_{mn}|^2 \ket$ when 
$\xi\ll 1$ and as a consequence Eq.~(\ref{F1}) underestimates the magnitude 
of $F$ for those $\xi$-values.  
Then, to fix this issue we propose the following expression
\begin{equation}
\label{F2}
F \approx C \xi^{-2} + 2 \ ,
\end{equation}
where $C$ is a fitting constant, to describe the curves $F$ vs.~$\xi$.
In Fig.~\ref{Fig8} we also show that Eq.~(\ref{F2}) fits reasonably well the
numerical data.

\section{Conclusions}

We study scattering and transport properties of tight-binding random 
networks characterized by the number of nodes $N$ and the average 
connectivity $\alpha$.  

We observed a smooth crossover from localized to delocalized behavior 
in the scattering and transport properties of the random 
network model by varying $\alpha$ from small ($\alpha\to 0$)
to large ($\alpha\to 1$) values. 
We show that all the scattering and transport quantities studied here 
are independent of $N$ once $\xi=\alpha N$ is fixed. 
Moreover, we proposes a heuristic and universal relation between 
the average scattering matrix elements $\bra |S_{mn}|^2 \ket$, the 
average conductance $\bra T \ket$, and the shot noise power $P$ 
and the disorder parameter $\xi$. See Eq.~(\ref{Xxi}).
As a consequence, we observed that the onset of the transition takes place at 
$\delta\xi\approx 0.1$; i.e., for $\delta\xi< 0.1$ the networks are in 
the insulating regime. While the onset of the Random Matrix Theory limit
is located at $\delta\xi\approx 10$; that is, for $\delta\xi>10$ the networks 
are in the metallic regime. Also, the metal-insulator transition point
is clearly located at $\delta\xi\approx 1$; see red dashed curves in 
Figs.~\ref{Fig6}(b) and \ref{Fig7}(b). Here, $\delta\in[0.2,0.4]$ is a 
parameter that 
slightly depends on the number of attached leads to the network but also 
on the quantity under study, see inserts of Figs.~\ref{Fig6}(b) and \ref{Fig7}(b).

Since our random network model is represented by an ensemble of sparse real 
symmetric random Hamiltonian matrices, in addition to random graphs of the
Erd\H{o}s-R\'enyi--type and complex networks, we expect our results to be also 
applicable to physical systems characterized by sparse Hamiltonian matrices, 
such as quantum chaotic and many-body systems.

\begin{acknowledgments}
This work was partially supported by VIEP-BUAP grant MEBJ-EXC13-I 
and PIFCA grant BUAP-CA-169.
\end{acknowledgments}


\begin{thebibliography}{99}
\bibitem{note1}
We remark that in the original Erd\H{o}s-R\'enyi random 
graph model the on-site potentials are set to zero while 
the hopping integrals are all equal to one.
\end{thebibliography}
\end{document}